# Attention Guided Metal Artifact Correction in MRI using Deep Neural Networks

Jee Won Kim, Kinam Kwon, Byungjai Kim, and HyunWook Park*
School of Electrical Engineering, Korea Advanced Institute of Science and Technology (KAIST)
Daejeon, South Korea
{jwkim, knkwon, byungjai}@athena.kaist.ac.kr, hwpark@kaist.ac.kr

**Abstract**

*An attention guided scheme for metal artifact correction in MRI using deep neural network is proposed in this paper. The inputs of the networks are two distorted images obtained with dual-polarity readout gradients. With MR image generation module and the additional data consistency loss to the previous work [1], the network is trained to estimate the frequency-shift map, off-resonance map, and attention map. The attention map helps to produce better distortion-corrected images by weighting on more relevant distortion-corrected images where two distortion-corrected images are produced with half of the frequency-shift maps. In this paper, we observed that in a real MRI environment, two distorted images obtained with opposite polarities of readout gradient showed artifacts in a different region. Therefore, we proved that using the attention map was important in that it reduced the residual ripple and pile-up artifacts near metallic implants.*

## 1. Introduction

Magnetic resonance imaging (MRI) provides anatomical and physiological information for clinical diagnosis. For those who get shoulder/hip/knee replacement surgeries, follow-up MRI imaging is required to check the progress of the surgery. However, the diagnosis of lesions near metal is problematic because of the large susceptibility of metallic implants causes severe distortion in images. Therefore, correction of metal artifacts is essential for clinical diagnosis.

Since metallic implants with large susceptibility cause wide ranges of off-resonance frequencies and cause severe distortion like blurring, signal misregistration, and signal voids. Several works have been developed to solve these problems. Some techniques estimated field inhomogeneity maps by acquiring two distorted images with dual-polarity spin-echo images and then used the maps to reduce the distortion in metallic object images numerically [2]. In other approaches, three-dimensional multispectral imaging (3D-MSI) methods acquired multiple spectral-bin images with different RF frequency offsets [3]. The geometric shifts are reduced significantly to a single-voxel level by this method, but residual pile-up and ripple induced by the off-resonance frequencies caused by the large local gradient still remain. Recently, Shi et al. reduced the residual pile-up and ripple artifacts in relatively short imaging time by estimating off-resonance frequency maps from spectral-bin images with two distorted images obtained with dual-polarity readout gradients [4].

The performance of deep neural networks is remarkable not only in image processing but also in medical imaging applications, such as classification, segmentation, reconstruction, and so on [5]. The deep neural networks in medical image reconstruction are superior in reducing reconstruction time and data-driven regularization for ill-posed problems, but sufficient training data are needed to avoid overfitting problems. Since collecting sufficient training data is difficult in medical imaging because of data privacy issues and the ambiguities of the ground truth images, several methods to combat this problem was to acquire simulation training data from natural images and developed unsupervised learning methods, which did not need ground truth [1].

In the previous work, unsupervised learning was applied to correct metal artifacts [1]. The architecture consisted of a deep neural network and MRI physics-based image generation module. In the training phase, the frequency-shift maps between two distorted images acquired using dual-polarity readout gradients are estimated by a U-net and MRI physics-based image generation module. The estimated frequency-shift maps are used to generate a distorted image from the opposite polarity distorted image. In the test phase, distortion-corrected images were obtained from the estimated frequency-shift maps and two distorted input images by the MRI physics-based image generation module. In the previous work, the distortion-corrected image was obtained by just averaging of the two outputs from two distorted images in the test phase [1]. In this work, we add an attention guided scheme. Then the proposed method adaptively weights on relevant outputs determined by data consistency loss to produce better distortion-corrected images. Recently, the attention mechanism is used in several works such as translation [6], image classification, [7] and so on, for helping neural networks to focus on most related features during training. Therefore, adding attention map to the previous work helps deep neural networks to focus more on the related output of the architecture to obtain the distortion-corrected image.

## 2. Theory

For this paper to be self-contained, we briefly review the theory of MR image signal modeling in existing of off-resonance frequencies [1]. In the presence of off-resonance caused by metallic implants, the MR image is more distorted along the frequency encoding direction than the phase encoding direction in the typical spatial encoding mechanism.

In the presence of off-resonance frequencies, $\delta v = \frac{\gamma}{2\pi}\delta B_0$, where $\delta B_0$ is the field inhomogeneity including magnetic susceptibility, the reconstructed MR image ($\tilde{I}$) with a readout gradient ($G_x$) can be written as follows:

$$\tilde{I}(\tilde{x}) = F_{G_x}(I, \delta v(x)) = \int \left(\int I(x) e^{-j2\pi k_x\left(x + \frac{2\pi \delta v(x)}{\gamma G_x}\right)} dx\right) e^{j2\pi \tilde{x} k_x} dk_x \quad (1)$$

where $F_{G_x}$ denotes an MR image generation operator. Here, $x$ and $k_x$ represent the image domain in the frequency-encoding direction, and the corresponding frequency domain, respectively. $I(x)$ denotes the average proton density at $x$ including $T_1$ and $T_2$ effects. The gyromagnetic ratio of the imaged nuclei is denoted as $\gamma$.

By using the MR image generation module and the Jacobian of the transformation [1], two distorted MR images ($I_+$ and $I_-$) acquired with dual-polarity readout gradients ($G_x$ and $-G_x$) can be represented as follows, respectively:

$$I_+(x') = F_{G_x}(I_0(x), \delta v(x)) = I_0(x)/\left|1 + \frac{2\pi}{\gamma G_x}\frac{\partial \delta v(x)}{\partial x}\right| \quad (2)$$

$$I_-(x'') = F_{-G_x}(I_0(x), \delta v(x))$$
$$= F_{G_x}(I_0(x), -\delta v(x)) = I_0(x)/\left|1 - \frac{2\pi}{\gamma G_x}\frac{\partial \delta v(x)}{\partial x}\right| \quad (3)$$

where $I_o$ is the corresponding distortion-free image, $x' = x + 2\pi\delta v(x)/\gamma G_x$, and $x'' = x - 2\pi\delta v(x)/\gamma G_x$.

## 3. Methods

### 3.1 Attention guided metal artifact correction framework

In the previous work, Kwon et al. showed that the distortion-free image can be obtained with half of the estimated frequency-shift maps ($\delta w_+/2$ and $\delta w_-/2$) between the two distorted images ($I_+$ and $I_-$) [1]. The frequency-shift maps and the intermediate distortion-corrected images ($\hat{I}_{+\to 0}$ and $\hat{I}_{-\to 0}$) can be estimated as follows:

$$\delta w_+(x') = arg\min_{\delta v}\left\|F_{G_x}(I_+(x'), \delta v(x')) - I_-(x')\right\|_2^2 \quad (4)$$

$$\delta w_-(x'') = arg\min_{\delta v}\left\|F_{G_x}(I_-(x''), \delta v(x'')) - I_+(x'')\right\|_2^2 \quad (5)$$

$$\hat{I}_{+\to 0} \cong F_{G_x}(I_+(x'), \delta w_+(x')/2) \quad (6)$$

$$\hat{I}_{-\to 0} \cong F_{G_x}(I_-(x''), \delta w_-(x'')/2). \quad (7)$$

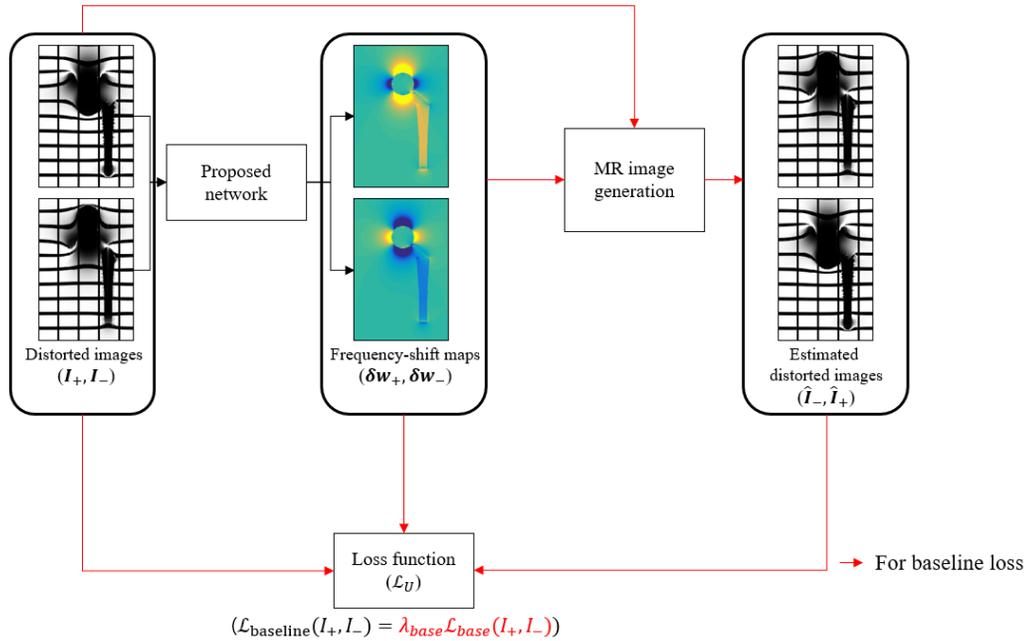

Figure 1: Scheme diagram of the baseline loss

In the part of the training baseline loss (Figure 1), base loss function ($\mathcal{L}_{base}$) which minimizes the mean squared error between the input distorted images ($I_+$ and $I_-$) and the estimated distorted images ($\hat{I}_+$ and $\hat{I}_-$) encourages networks to estimate the frequency-shift maps according to Equations (4) and (5). The $\mathcal{L}_{base}$ is defined as follows:

$$\mathcal{L}_{base}(I_+, I_-) = \left\|\hat{I}_+ - I_+\right\|_2^2 + \left\|\hat{I}_- - I_-\right\|_2^2. \quad (8)$$

In the test phase (dashed red line in Figure 2), the distortion-corrected image is obtained with the attention map ($\rho$) and half of the estimated frequency-shift maps ($\delta w_+/2$ and $\delta w_-/2$) as follows:

$$\hat{I}_0 \cong \rho * \hat{I}_{+\to 0} + (1-\rho) * \hat{I}_{-\to 0} \quad (9)$$

With the attention map ($\rho$), the more relevant information between the intermediate results from Equations (6) and (7) are weighted for acquiring the distortion-corrected image. The attention map ($\rho$) is defined with sigmoid function which returns a value between 0 and 1. In a real MRI environment, the influence of the off-resonance frequencies can be slightly different by eddy currents and time-varying fields, and so on. Therefore, it is important to use the attention map in the real MRI experiments. The details of the attention guided scheme will be discussed in the following section.

### 3.2 Data consistency loss for attention map

The attention map determines which of the two distortion-corrected images has more relevance for each pixel. Thus, to make the neural network automatically decide the relevance, the data consistency loss is considered in training as an unsupervised manner. With the final distortion-corrected image ($\hat{I}_0$) and corresponding off-resonance maps ($\delta v(x)$), input images can be estimated as follows:

$$\hat{I}_{0\to +}(x) = F_{G_x}\left(\hat{I}_0(x), \delta v(x)\right) \quad (10)$$
$$\hat{I}_{0\to -}(x) = F_{-G_x}\left(\hat{I}_0(x), \delta v(x)\right) \quad (11)$$

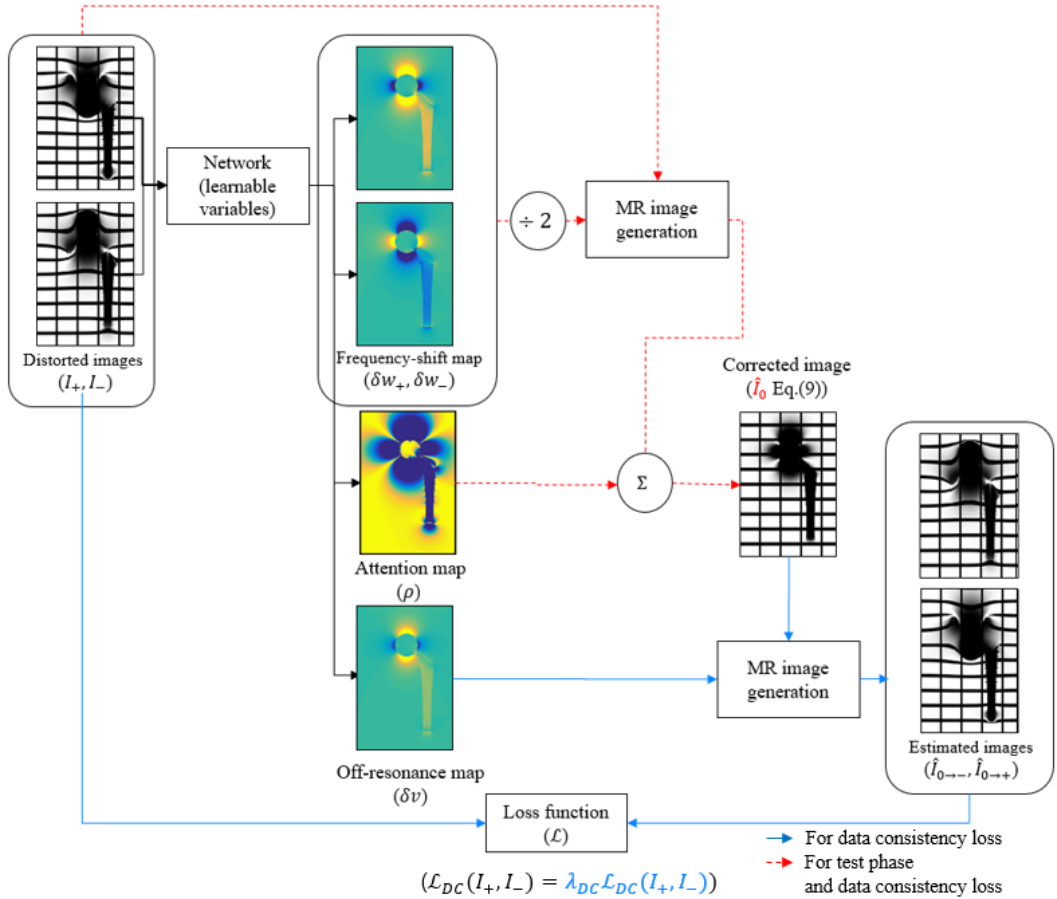

Figure 2: Scheme diagram of the data consistency loss

where $\hat{I}_{0\to+}$ is a distorted image acquired with a positive polarity readout gradient from $\hat{I}_0$ and $\hat{I}_{0\to-}$ is a distorted image acquired with a negative polarity readout gradient from $\hat{I}_0$. Then, the data consistency loss ($\mathcal{L}_{DC}$) is defined as follows:

$$\mathcal{L}_{DC}(I_+, I_-) = \|\hat{I}_{0\to+} - I_+\|_2^2 + \|\hat{I}_{0\to-} - I_-\|_2^2 \quad (12)$$

By using $\mathcal{L}_{DC}$ in the part of the training phase (blue and dashed red arrows in Figure 2), the attention map ($\rho$) is trained to adaptively choose the weights depending on the relevance of the outputs for the distortion-corrected image.

### 3.3 Total Loss function

In the total loss function ($\mathcal{L}_{total}$), to encourage smooth frequency-shift maps and off-resonance map, total variation loss is included, respectively. The total loss is given as follows:

$$\mathcal{L}_{total}(I_+, I_-) = \lambda_{base}\mathcal{L}_{base}(I_+, I_-) + \lambda_{DC}\mathcal{L}_{DC}(I_+, I_-) + \lambda_{TV}(\mathcal{L}_{TV}(\delta v) + \mathcal{L}_{TV}(\delta w)) \quad (13)$$

where $\lambda_*$ are regularization parameters for loss functions. The deep neural network is trained to minimize the loss function. Each regularization parameter is chosen empirically as $\lambda_{base} = 1$, $\lambda_{DC} = 1$, $\lambda_{TV} = 10^{-4}$.

### 3.4 Implementation details of the Neural Networks and Datasets

In this work, as shown in Figure 3, a modified version of the U-net [8] is used for the training. To consider the local perturbations induced by the metallic implant, convolutional layers with a kernel size of 3×3 are used [8]. The zero-padded inputs of the convolutional layers are used for keeping the spatial dimensions. After each convolutional layer, batch normalization (BN) and the rectified linear unit (ReLU) are sequentially applied (red arrows). To increase the receptive field for considering the dispersion of the metal-induced artifacts, a multi-scale architecture with down-sampling and up-sampling is used. In down-sampling, max pooling layers with a kernel size of 3×3 and a stride of 2×2 are utilized. In up-sampling, the transposed convolution operation with a kernel size of 3×3 and a stride of 2×2 is used and indicated as 'up-conv 2×2' in Figure 3. The number of channels is written beneath the layers. To avoid information loss, a concatenate layer for each scale (yellow arrows) is utilized. The skip layer (gray arrows) helps the learning process to be more stable and efficient [9]. The weight values are initialized with uniform distribution. The deep learning package, 'TensorFlow', with an Intel Xeon (E5-2687) CPU and an NVDIA TITNA-x GPU is used [10]. The Adam optimizer with a learning rate of $10^{-4}$ is used for minimizing the loss function. The normalized training datasets with the maximum value of each dataset are randomly cropped into 160×160 [11]. In experiments, 100,000 synthesized data with additional augmentation are used for the training.

In clinical applications, it is unrealistic to collect distortion-free MR data from imaging regions near metallic implants because of unpractically long imaging time for obtaining in-vivo images. Therefore, synthetic data are utilized to train the proposed network and compared networks. The synthetic data are generated to mimic distorted and distortion-free MR images in the presence of off-resonance frequencies [11].

### 3.5 MRI Experiments

We acquired phantom test set with the 3T MRI system (Siemens Verio, Germany). For phantom, we used a hip joint replacement implant consisting of a femoral head and a femoral stem. The implant was dipped in CuSO4 solution with the fixation of the acryl grids. We used non-slice-selective three-dimensional fast spin-echo sequence and RF frequency offsets were spaced by 1kHz from -11.5kHz to 11.5kHz for multiple spectral-bin images. The imaging parameters are as follows: TR/TE = 3500/8ms, the number of spectral bins = 24 (Gaussian RF profiles with FWHM = 2.25 kHz for spectral selection), matrix size = $256 \times 128 \times 20$, voxel size = $1 \times 1 \times 5$ mm$^3$, echo-train length = 16, and readout bandwidth = 780 Hz/pixel.

For comparison, the phantom image is reconstructed with two different methods. The first method is based on the multi-acquisition with variable resonance image combination (MAVRIC) [3]. MAVRIC (+) and MAVRIC (-) are obtained by combining the multi-spectral images acquired with positive and negative-polarity readout gradients, respectively. The other method acquires a distortion-free image by using MR generation module and half of the estimated frequency-shift maps ($\delta w_+/2$ and $\delta w_-/2$) [1]. As explained in Equations (6) and (7), $\hat{I}_{+\to 0}$ is obtained as an intermediate distortion-corrected image with $\delta w_+/2$ and $\hat{I}_{-\to 0}$ is obtained as an intermediate distortion-corrected image with $\delta w_-/2$. For the second method and the proposed method, every single spectral-bin image is corrected by the network and then multiple spectral-bin images are combined by using root-sum-of-square (RSOS) along the spectral-bin dimension.

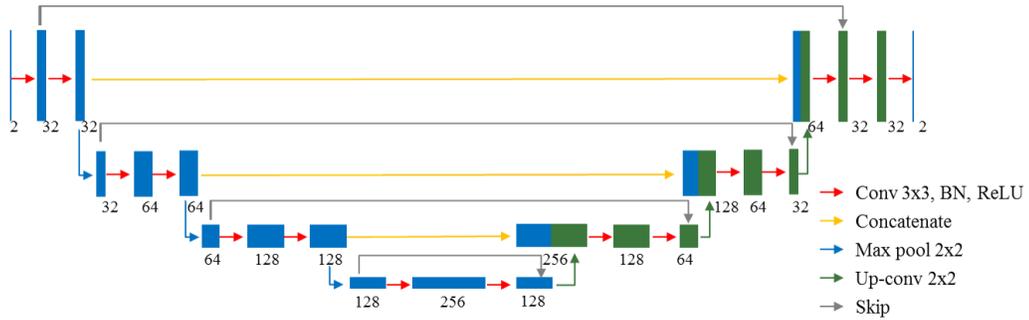

Figure 3: Scheme diagram of the deep neural networks.

## 4. Results

Figure 4 shows the distortion-corrected images of the phantom of metallic implants reconstructed by several methods. As shown in Figure 4, since the bulk shifts reduce to a single-voxel level, the blurring artifacts are minimized in both MAVRIC(+) and MAVRIC(-). However, the residual ripple and pile-up artifacts still significantly degrade the images (blue arrows). In the intermediate distortion-corrected images ($\hat{I}_{+\to 0}$ and $\hat{I}_{-\to 0}$), even though the artifacts caused by severe off-resonance frequencies near the metallic implants are reduced compared to MAVRIC method, we could still observe the ripple and pile-up artifacts (blue arrows). The results of the proposed method show significant improvements compared to other methods. The residual ripple artifacts nearby metallic implants nearly disappeared.

Figure 5 shows the outputs of the neural networks in the test phase. From the left, $\delta w_+$ and $\delta w_-$ are the estimated frequency-shift maps, $\delta v$ is the estimated off-resonance map and $\rho$ is the attention map. The range of estimated frequency-shift maps and the estimated off-resonance map is from -6 to 6 and the range of the attention map is from 0 to 1. In the test phase, half of the estimated frequency-shift maps ($\delta w_+/2$ and $\delta w_-/2$) and the attention map are used for acquiring the distortion-corrected image.

## 5. Discussions

The proposed method used two distorted images obtained with dual-polarity readout gradients as inputs of the deep neural networks. Subsequently, we acquired the distortion-corrected image considering the MRI environment by using deep neural network, MR image generation module, and attention map. Theoretically, in an ideal MRI environment without the off-resonance frequencies, even though the readout gradient polarity is different, the results of the MR image should be the same. However, in a real MRI environment, when we acquire the MR image with different readout gradient polarity, the acquired MR image is different because of the off-resonance frequencies. We could observe this effect in Figure 4. The regions of the artifacts are different in MAVRIC(+) and MAVRIC(-).

In addition, eddy-current and field-varying are also included in the real MRI experiments. Without these effects, the results of two intermediate distortion-corrected images ($\hat{I}_{+\to 0}$ and $\hat{I}_{-\to 0}$) will be the same. However, as shown in Figure 4, the artifacts occurred in different regions. Therefore, using the attention map is important. This helps networks to consider a real MRI environment and focus more on relevant information of two intermediate distortion-corrected images ($\hat{I}_{+\to 0}$ and $\hat{I}_{-\to 0}$) for obtaining the distortion-corrected image ($\hat{I}_0$). As shown in Figure 4, with this attention map, we acquired an improved distortion-corrected image ($\hat{I}_0$) compared to other methods.

As shown in the attention map ($\rho$) in Figure 5, the regions of ripple artifacts in the distortion-corrected map in Figure 4 have low attention values (red arrows). We conclude that the attention map is well learned with deep neural networks and the proposed method follows the existing methods which estimate off-resonance map for metal artifact correction [2, 4] because we could observe that attention value was high where distortion doesn't exist and attention value was low where distortion like ripples exists. Therefore, the proposed method gives explainability compared to the Kwon et al. method [1]. The proposed method was carried out only with phantom images. In-vivo experiments will be further studied. In addition, the quantitative comparison with the results of the previous work [1] is required in future works.

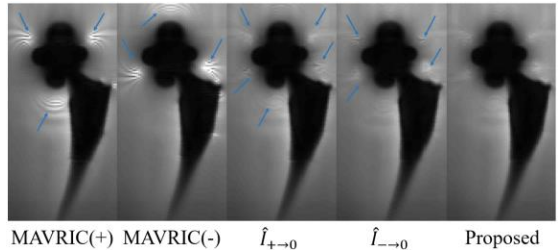

MAVRIC(+)  MAVRIC(-)  $\hat{I}_{+\to 0}$  $\hat{I}_{-\to 0}$  Proposed

Figure 4: Phantom images reconstructed with several methods including the proposed method.

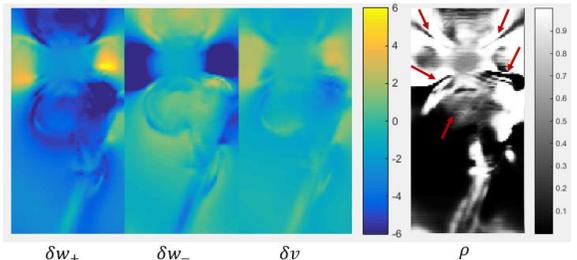

$\delta w_+$  $\delta w_-$  $\delta v$  $\rho$

Figure 5: The outputs from the neural networks in the test phase. From the left, $\delta w_+$ and $\delta w_-$ are the estimated frequency-shift maps, $\delta v$ is the estimated off-resonance map and $\rho$ is the attention map.